\documentclass[twoside,english,tightenlines]{iopart}
\usepackage[T1]{fontenc}
\usepackage[latin9]{inputenc}
\usepackage{geometry}
\usepackage{babel}
\usepackage{amsbsy}
\usepackage{amssymb}
\usepackage{graphicx}
\usepackage[numbers]{natbib}
\usepackage[unicode=true,pdfusetitle,
 bookmarks=true,bookmarksnumbered=false,bookmarksopen=false,
 breaklinks=false,pdfborder={0 0 1},backref=section,colorlinks=false]
 {hyperref}

\makeatletter
\usepackage{iopams}
\usepackage{setstack}

\newcommand{\1}{\leavevmode{\rm 1\ifmmode\mkern  -4.8mu\else\kern -.3em\fi I}}

\usepackage{babel}
\usepackage{mciteplus}

\makeatother

\begin{document}

\title{Probability density of quantum expectation values}

\author{L Campos Venuti and P Zanardi }

\address{Department of Physics and Astronomy and Center for Quantum Information
Science \& Technology, University of Southern California, Los Angeles,
California 90089-0484, USA }

\ead{lcamposv@usc.edu }
\begin{abstract}
We consider the quantum expectation value $\mathcal{A}=\langle\psi|A|\psi\rangle$
of an observable $A$ over the state $|\psi\rangle$. We derive the
exact probability distribution of $\mathcal{A}$ seen as a random
variable when $|\psi\rangle$ varies over the set of all pure states
equipped with the Haar-induced measure. The probability density is
obtained with elementary means by computing its characteristic function,
both for non-degenerate and degenerate observables. To illustrate
our results we compare the exact predictions for few concrete examples
with the concentration bounds obtained using Levy's lemma. Finally
we comment on the relevance of the central limit theorem and draw
some results on an alternative statistical mechanics based on the
uniform measure on the energy shell. 
\end{abstract}
\maketitle

\section{Introduction}

The role of probability distributions in quantum theory cannot be
overestimated. Arguably the most important of those distributions
is the one describing the statistics of possible outcomes of the measurement
of the observable associated with the self-adjoint operator $A$ while
the system is in the pure state $|\psi\rangle$: $P_{A}\left(a\right):=\langle\psi|\delta\left(A-a\right)|\psi\rangle.$
This function is supported on the numerical range of $A$ and, for
bounded $A$ can be equivalently characterized by the set of its moments
(see e.g.~\citep{tao_topics_2012}): $m_{k}:=\langle\psi|A^{k}|\psi\rangle\,(k\in{\bf N}),$
i.e., the expectation values of the family of observables $\{A^{k}\}_{k\in{\bf N}}$
in the state $|\psi\rangle.$ This latter, quite often, can be itself
regarded as a random variable distributed according to some prior
density that depends on the problem under consideration. For example,
in the context of equilibration dynamics of closed quantum systems
\citep{linden_quantum_2009,campos_venuti_exact_2011} one is interested
in the quantity $a(t):=\langle\psi(t)|A|\psi(t)\rangle$, where $|\psi(t)\rangle:=e^{-itH}|\psi\rangle$
and $H$ is the Hamiltonian operator of the system. If one monitors
$A$ by sampling time instants uniformly over a the interval $\left[0,T\right]$
 the underlying probability space for the $|\psi\left(t\right)\rangle$'s
is the segment $\left[0,T\right]$ equipped with the uniform measure
$dt/T$. In this case averaging over the quantum states amounts to
perform the time average $1/T\int_{0}^{T}a(t)dt.$

Another possibility that recently gained relevance for the foundation
of statistical mechanics \citep{popescu_entanglement_2006} is to
consider $\langle\psi|A|\psi\rangle$ and let $|\psi\rangle$ vary
over the full unit sphere of the, say $d$-dimensional, Hilbert space
space. This manifold is transitively acted upon by the group of all
$d\times d$ unitary matrices $\mathbb{U}\left(d\right)$ and therefore
inherits a natural invariant measure from the unique group-theoretic
invariant measure over $\mathbb{U}\left(d\right)$ i.e., the Haar
measure.

In this paper we will address precisely this latter setting and compute
the probability distribution for the quantum expectations $\langle\psi|A^{k}|\psi\rangle$
seen as random variables over the unit sphere of the Hilbert space
equipped with the measure induced by the Haar measure. In this way
the function $P_{A}$ itself becomes a sort probability-density valued
random variable that can be partially characterized by the probability
densities of its moments $m_{k}$ over the unit sphere. We will show
that these probability densities can be determined with elementary
tools and explicit analytical expressions for their characteristic
functions can be obtained. To be specific we will concentrate on the
probability density defined as $P_{\mathcal{A}}(a):=\overline{\delta\left(\langle\psi|A|\psi\rangle-a\right)}$
where the overline $\overline{f\left(\psi\right)}=\int D\psi\, f\left(\psi\right)$
indicates Haar-induced averages over pure states. The probability
distribution for $\langle\psi|A^{k}|\psi\rangle$ is trivially obtained
with the substitution $A\to A^{k}$. 

The probability density $P_{\mathcal{A}}\left(a\right)$ has been
first considered in a series of works \citep{brody_quantum_1998,brody_microcanonical_2005,bender_solvable_2005,brody_quantum_2007,brody_quantum_2007-1}
which introduced the so called {}``quantum microcanonical  equilibrium''
(QME), an alternative statistical mechanics based on a generalization
of the postulate of equal a-priori probability. This postulate states
that, at equilibrium, all the energy eigenstates in a given energy
shell are equally probable and it leads to the familiar microcanical
equilibrium state $\rho_{MC}=\Pi_{E}/d$ where $\Pi_{E}$ is the projection
onto the space of energies between $E$ and $E+\Delta$, {[}$d$ is
the dimension of its range and $\Delta$ is a small parameter (see
e.g.~\citep{huang_statistical_1963}){]}. In the QME setting instead
one also allows for quantum superposition of energy eigenstates. This
leads one to consider all normalized states in the Hilbert space consistent
with a given energy taken with uniform probability, that is, the Haar
induced measure over the pure states. We will come back to QME in
Sec.~\ref{sec:A-note-on}. In particular we will answer the question
whether QME gives rise to extensive free energy and the typical size
of fluctuations. 

The probability density $P_{\mathcal{A}}(a)$ has been considered
also in a series of recent works \citep{dunkl_numerical_2011,dunkl_numerical_2011-1,puchala_restricted_2012}
(see also \citep{gallay_numerical_2012} for an entry into the mathematical
literature) where $P_{\mathcal{A}}(a)$ is sometimes referred to as
numerical shadow. In the present article we give extra care to the
physical case where the total Hilbert space is that of a many-body
systems and observables are extensive operators. Moreover, for the
first time, we give explicit formulae for the moment generating function
and the probability density for the general case where $A$ is not
necessarily a non-degenerate operator.

\section{Preliminaries}

Our key object is $\mathcal{A}\left(\psi\right)=\langle\psi|A|\psi\rangle.$
Computing the first few moments of $\mathcal{A}\left(\psi\right)$
is a relatively easy task. The first moment reads 
\begin{equation}
m_{1}=\overline{\mathcal{A}}=\int D\psi\,\tr\left(A|\psi\rangle\langle\psi|\right)=\frac{\tr\left(A\right)}{d}\,.\label{eq:average}
\end{equation}
 a result which follows from $\overline{|\psi\rangle\langle\psi|}=\1/d$
\citep{weyl_classical_1997}. A closed formula for the general moment
can be obtained by noting that 
\begin{equation}
\overline{|\psi\rangle\langle\psi|^{\otimes n}}=\frac{1}{\left(\begin{array}{c}
d+n-1\\
n
\end{array}\right)}\frac{1}{n!}\sum_{\pi\in S_{n}}P_{\pi}\,.\label{eq:general_average}
\end{equation}
 Here $P_{\pi}$ is the operator that enacts the permutation $\pi$
in $\mathcal{H}^{\otimes n}$ and $S_{n}$ is the symmetric group
of $n$ elements. For a proof of (\ref{eq:general_average}) see e.g.~\citep{goodman_representations_2000}.
The proportionality constant is obtained noting that $\left(\begin{array}{c}
d+n-1\\
n
\end{array}\right)$ is the dimension of the totally symmetric space, and the remaining
operator is an orthogonal projector. Using eq.~(\ref{eq:general_average})
one obtains the following closed expression for the $n$-th moment

\begin{eqnarray}
m_{n} & = & \frac{\left(d-1\right)!}{\left(d+n-1\right)!}\sum_{\pi\in S_{n}}\tr\left(P_{\pi}A^{\otimes n}\right)\nonumber \\
 & = & \frac{\left(d-1\right)!}{\left(d+n-1\right)!}\sum_{\pi\in S_{n}}\sum_{\sigma_{1}=1}^{d}\cdots\sum_{\sigma_{n}=1}^{d}A_{\sigma_{1},\sigma_{\pi\left(1\right)}}\cdots A_{\sigma_{n},\sigma_{\pi\left(n\right)}}\,.\label{eq:general_moment}
\end{eqnarray}
 This expression can be thought of as a sum of contractions and represents
a sum of products of traces of $A$. For instance for $n=2,3$ one
has 
\begin{equation}
m_{2}=\frac{\tr\left(A\right)^{2}+\tr\left(A^{2}\right)}{d\left(d+1\right)},\qquad m_{3}=\frac{\tr\left(A\right)^{3}+3\tr\left(A^{2}\right)\tr\left(A\right)+2\tr\left(A^{3}\right)}{d(d+1)(d+2)}\,.\label{eq:m2-m3}
\end{equation}
With a little extra work one gets for $n=4$
\begin{equation}
m_{4}=\frac{\tr\left(A\right)^{4}+6\left[\tr\left(A^{2}\right)\right]\left[\tr\left(A\right)\right]^{2}+3\left[\tr\left(A^{2}\right)\right]^{2}+8\tr\left(A\right)\tr\left(A^{3}\right)+6\tr\left(A^{4}\right)}{d(d+1)(d+2)(d+3)}.\label{eq:m4}
\end{equation}
Although the moments can be obtained in closed form it seems difficult
to obtain the probability density following this approach. Instead
our procedure will be that of computing directly the characteristic
function $\chi\left(\lambda\right):=\overline{e^{i\lambda\mathcal{A}\left(\psi\right)}}$
and obtain the probability density via Fourier transform.

Choosing a basis $|j\rangle$, ($j=1,\ldots,d$) and calling $z_{j}=\langle j|\psi\rangle$
we observe that we can write the average over $|\psi\rangle$ as 
\begin{equation}
\overline{f\left(\psi\right)}=C\int\delta\left(\sum_{j=1}^{d}\left|z_{j}\right|^{2}-1\right)f\left(\psi\right)d^{2}\boldsymbol{z}\label{eq:ave}
\end{equation}
 where we defined 
\begin{equation}
\int d^{2}\boldsymbol{z}=\prod_{i=1}^{d}\int_{-\infty}^{\infty}\int_{-\infty}^{\infty}\frac{dx_{i}dy_{i}}{\pi}.
\end{equation}
 The normalization constant $C$ can be computed with the same technique
that we are going to show and it turns out to be equal to $\left(d-1\right)!$.
Using the Fourier representation for the delta function in eq.~(\ref{eq:ave})
we obtain a Gaussian integral that can be computed. For simplicity
we use the basis that diagonalizes $A$: $A=\sum_{j}a_{j}|j\rangle\langle j|$.
Calling $D_{A}=\mathrm{diag}\left\{ a_{1},a_{2},\ldots,a_{d}\right\} $
we can write the characteristic function as 
\begin{equation}
\chi\left(\lambda\right)=\left(d-1\right)!\int d^{2}\boldsymbol{z}\int_{-\infty}^{\infty}\frac{dr}{2\pi}e^{ir\left(z^{\dagger}z-1\right)}e^{i\lambda z^{\dagger}D_{A}z-\epsilon z^{\dagger}z}\,.
\end{equation}
 As customary we introduced a small positive $\epsilon$ in order
to make the Gaussian integral absolutely convergent. The Gaussian
integration gives 
\begin{equation}
\chi\left(\lambda\right)=\frac{\left(d-1\right)!}{\left(-i\right)^{d}}\int_{-\infty}^{\infty}\frac{dr}{2\pi}\frac{e^{-ir}}{\prod_{j}\left(r-r_{j}\right)},\quad r_{j}=-\lambda a_{j}-i\epsilon\label{eq:chi-pre}
\end{equation}
 At this point we make the important assumption that all eigenvalues
of $A$ are non-degenerate. We will treat the general case in section
\ref{sec:General-case}. Under these conditions the integrand in eq.~(\ref{eq:chi-pre})
has only simple poles and the integral is easily evaluated with residues
closing the circle in the in the lower half-plane. The result is,
after sending $\epsilon\to0$, 
\begin{equation}
\chi\left(\lambda\right)=\frac{\left(d-1\right)!}{\left(i\lambda\right)^{d-1}}\sum_{k=1}^{d}\frac{e^{i\lambda a_{k}}}{\prod_{j\neq k}\left(a_{k}-a_{j}\right)}\,.\label{eq:chi}
\end{equation}
 Although it might not be readily apparent from eq.~(\ref{eq:chi}),
$\chi\left(\lambda\right)$ is actually regular in $\lambda=0$. This
fact follows from a set of identities proven in \ref{sec:Some-useful-identities}
stating that 
\begin{equation}
\sum_{k=1}^{d}\frac{\left(a_{k}\right)^{n}}{\prod_{j\neq k}\left(a_{k}-a_{j}\right)}=\left\{ \begin{array}{cl}
0 & 0\le n\le d-2\\
1 & n=d-1
\end{array}\right.\,.\label{eq:identity}
\end{equation}
 Applying eq.~(\ref{eq:identity}) to eq.~(\ref{eq:chi}) we thus
see that $\chi\left(\lambda\right)$ is regular at $\lambda=0$ and
being a linear combination of analytic functions it is in fact analytic
in the whole complex plane (entire). A simple way to remember that
$\chi\left(\lambda\right)$ must be regular at $\lambda=0$ is to
note that $\chi\left(\lambda\right)=1+m_{1}\left(i\lambda\right)+O\left(\lambda^{2}\right)$.
In fact this very same approach can be used to prove eq.~(\ref{eq:identity}).
Using eq.~(\ref{eq:identity}) and eq.~(\ref{eq:chi}) we readily
obtain the Taylor series of $\chi\left(\lambda\right)$ 
\begin{eqnarray}
\chi\left(\lambda\right) & = & \sum_{n=0}^{\infty}m_{n}\frac{\left(i\lambda\right)^{n}}{n!}\label{eq:chi-series}\\
m_{n} & = & \frac{1}{\left(\begin{array}{c}
n+d-1\\
n
\end{array}\right)}\sum_{k=1}^{d}\frac{\left(a_{k}\right)^{n+d-1}}{\prod_{j\neq k}\left(a_{k}-a_{j}\right)}\,.\label{eq:moments_compact}
\end{eqnarray}
 Equation (\ref{eq:moments_compact}) is a quite compact expression
in place of the complicated eq.~(\ref{eq:general_moment}). Equating
eq.~(\ref{eq:moments_compact}) with eq.~(\ref{eq:general_moment})
we obtain the following non-trivial set of matrix identities valid
when the spectrum of $A$ is non-degenerate 
\begin{equation}
\frac{1}{n!}\sum_{\pi\in S_{n}}\tr\left(P_{\pi}A^{\otimes n}\right)=\sum_{k=1}^{d}\frac{\left(a_{k}\right)^{n+d-1}}{\prod_{j\neq k}\left(a_{k}-a_{j}\right)}\,.
\end{equation}

Since $\chi\left(\lambda\right)$ is entire, analytic continuation
is trivial and the moment generating function $M\left(y\right):=\overline{e^{y\mathcal{A}}}$
can be simply obtained by setting $i\lambda=y$ in eq.~(\ref{eq:chi}).

To obtain the probability density we must compute the Fourier transform
of $\chi$ 
\begin{equation}
P_{\mathcal{A}}\left(x\right)=\int_{-\infty}^{\infty}\frac{d\lambda}{2\pi}e^{-ix\lambda}\chi\left(\lambda\right).\label{eq:Fourier:integral}
\end{equation}
 We make use of the following trick. Since $\chi\left(\lambda\right)$
is well behaved in zero and decays at infinity, it belongs to the
space of tempered distribution $\mathcal{S}'\left(\mathbb{R}\right)$.
The Fourier transform is well defined for such functions and the result
is again a tempered distribution. Actually for $d\ge3$, $\chi$ is
also summable and so its Fourier transform can be safely defined by
the absolutely converging integral eq.~(\ref{eq:Fourier:integral}).
We proceed than forgetting about the behavior in $\lambda=0$ and
compute the Fourier transform as a linear combination of Fourier transform
of the functions $e^{i\lambda a_{k}}/\left(i\lambda\right)^{d-1}$.
Such functions however are not well behaved in zero (they are not
distributions) and must be regularized in order to compute the Fourier
transform. Since the total Fourier transform is well defined, the
result cannot depend on the regularization. A convenient regularization
of $1/\lambda^{n}$ for $n$ integer, is given in terms of the $n$-th
derivative of $\ln\left|\lambda\right|$, more precisely $\left.\lambda^{-n}\right|_{\mathrm{regularized}}:=\left(-1\right)^{n-1}\partial_{\lambda}\ln\left|\lambda\right|/\left(n-1\right)!$.
With this definition the Fourier transform, in the sense of distribution,
is 
\begin{equation}
\int_{-\infty}^{\infty}\frac{d\lambda}{2\pi}\frac{e^{i\lambda y}}{\left(i\lambda\right)^{n}}=\frac{y^{n-1}\mathrm{sign}\left(y\right)}{2\left(n-1\right)!}.
\end{equation}
 Using this equation and eq.~(\ref{eq:chi}) we obtain 
\begin{equation}
P_{\mathcal{A}}\left(x\right)=\frac{\left(d-1\right)}{2}\sum_{k}\frac{\left(a_{k}-x\right)^{d-2}\mathrm{sign}\left(a_{k}-x\right)}{\prod_{j\neq k}\left(a_{k}-a_{j}\right)}.\label{eq:P-density}
\end{equation}

By construction, $P_{\mathcal{A}}\left(x\right)$ has all the properties
of a probability density, moreover, since the probability of obtaining
a value outside the numerical range of $A$ must be zero, $P_{\mathcal{A}}\left(x\right)$
must be supported in $\left[\min\{a_{k}\},\max\{a_{k}\}\right]$ which,
once again, is not readily apparent from eq.~(\ref{eq:P-density}).
That $P_{\mathcal{A}}\left(x\right)$ is compactly supported follows
from Paley-Wiener theorem since it is the Fourier transform of an
$L^{2}\left(\mathbb{R}\right)$ analytic function. In any case a direct
proof that $P_{\mathcal{A}}\left(x\right)=0$ for $x$ outside $\left[\min\{a_{k}\},\max\{a_{k}\}\right]$
is given in \ref{sec:Some-useful-identities} and follows directly
form eq.~(\ref{eq:nano}).

\begin{figure}
\noindent \begin{centering}
\includegraphics[width=8cm]{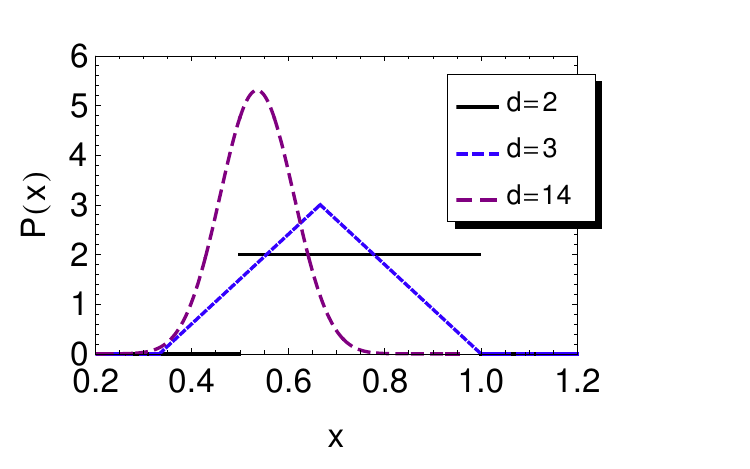} 
\par\end{centering}

\noindent \caption{Plot of the probability density eq.~(\ref{eq:P-density}) for various
$d$. In this particular case we set $a_{k}=k/d$. \label{fig:Pdf}}
\end{figure}

Generically $P_{\mathcal{A}}\left(x\right)$ is a bell-shaped curve
supported in $\left[\min\{a_{k}\},\max\{a_{k}\}\right]$ as can be
see from fig.~\ref{fig:Pdf}. The smoothness properties of $P_{\mathcal{A}}\left(x\right)$
can be understood noting that $\partial_{x}^{\left(d-2\right)}P_{\mathcal{A}}\left(x\right)$
is a piecewise constant function (a part from a bunch of delta peaks).

\subsection{Cumulative distribution function }

Having computed the probability density, the characteristic, and the
moment generating function, we would like now to compute the cumulative
distribution function (CDF) to complete the picture. The CDF can be
obtained via the following integral 
\begin{equation}
\mathrm{cdf}\left(x\right)=\mathrm{Prob}\left(\mathcal{A}\le x\right)=\int_{-\Omega}^{x}P_{\mathcal{A}}\left(t\right)dt\,,
\end{equation}
 as long as $-\Omega<\min\{a_{k}\}$. The integration of each term
in eq.~(\ref{eq:P-density}) gives 
\begin{equation}
\int_{-\Omega}^{x}\left(t-a_{k}\right)^{d-2}\mathrm{sign}\left(t-a_{k}\right)dt=\mathrm{sign}\left(x-a_{k}\right)\frac{\left(x-a_{k}\right)^{d-1}}{\left(d-1\right)}+\frac{\left(-\Omega-a_{k}\right)^{d-1}}{\left(d-1\right)}\,.
\end{equation}
 Clearly the result cannot depend on $\Omega$ as long as $-\Omega<\min\{a_{k}\}$.
In fact, using again formula (\ref{eq:nano}) of \ref{sec:Some-useful-identities},
the sum of all the terms containing $\Omega$ gives 
\begin{equation}
\frac{1}{2}\sum_{k=1}^{d}\frac{\left(-\Omega-a_{k}\right)^{d-1}}{\prod_{j\neq k}\left(a_{j}-a_{k}\right)}=\frac{1}{2}\,.\label{eq:nano2}
\end{equation}
 All in all the final expression for the cumulative is 
\begin{equation}
\mathrm{cdf}\left(x\right)=\frac{1}{2}+\frac{1}{2}\sum_{k}\mathrm{sign}\left(x-a_{k}\right)\frac{\left(x-a_{k}\right)^{d-1}}{\prod_{j\neq k}\left(a_{j}-a_{k}\right)}\label{eq:cdf}
\end{equation}

\section{General case\label{sec:General-case}}

So far we assumed that the spectrum of $A$ was non-degenerate. Although
this is the most common situation, one may want to consider the case
where $A$ is a projector. We then turn to the fully general case
where the spectrum of $A$ consists of $\ell$ distinct eigenvalues
$\{a_{j}\}_{j=1}^{\ell}$ with degeneracy $n_{j}$ satisfying $\sum_{j=1}^{\ell}n_{j}=d$.
For a change, let us compute the moment generating function $M\left(y,\Omega\right):=\overline{e^{y\left(\mathcal{A}-\Omega\right)}}$
instead of the characteristic function. The shift constant is chosen
to satisfy $\Omega>\max\{a_{j}\}$ so that for $y>0$ the Gaussian
integral is well defined. Gaussian integration gives the same result
as for the non-degenerate case, which we now re-write as 
\begin{equation}
M\left(y,\Omega\right)=\frac{\left(d-1\right)!}{\left(-i\right)^{d}}\int_{-\infty}^{\infty}\frac{dr}{2\pi}\frac{e^{-ir}}{\prod_{j=1}^{\ell}\left(r-r_{j}\right)^{n_{j}}}\,.\label{eq:integral_degenerate}
\end{equation}
 To compute the above integral with the residues we must compute the
following derivatives 
\begin{equation}
\mathrm{Res}\left[\frac{e^{-ir}}{\prod_{j=1}^{\ell}\left(r-r_{j}\right)^{n_{j}}},r=r_{k}\right]=\frac{1}{\left(n_{k}-1\right)!}\partial_{r}^{\left(n_{k}-1\right)}\left.\left[\frac{e^{-ir}}{\prod_{j\neq k}^{\ell}\left(r-r_{j}\right)^{n_{j}}}\right]\right|_{r=r_{k}}\label{eq:residue}
\end{equation}
 Alternatively, another possibility is to use the Chinese Remainder
Theorem (see e.g.~\citep{gaillard_around_2008}) %
\footnote{LCV whishes to thank Pierre-Yves Gaillard for pointing out this connection.%
} to get the partial fraction decomposition of the function in Eq.~(\ref{eq:integral_degenerate}).
This way one expresses the integrand in Eq.~(\ref{eq:integral_degenerate})
as a sum of functions with simple poles so that the residues can be
evaluated straightforwardly. The results coincide and in fact eq.~(\ref{eq:residue})
provides a way to obtain the partial fraction decomposition of the
function in Eq.~(\ref{eq:integral_degenerate}) by differentiation.
The procedure is outlined in \ref{sec:Degenerate-case}, and the final
result for the moment generating function is

\begin{eqnarray}
M\left(y,\Omega\right) & = & \left(d-1\right)!\sum_{k=1}^{\ell}\sum_{M_{k}=0}^{n_{k}-1}\frac{\left(-1\right)^{M_{k}}e^{y\left(a_{k}-\Omega\right)}}{y^{d+M_{k}-n_{k}}\left(n_{k}-1-M_{k}\right)!}\times\nonumber \\
 &  & \times\sum_{\sum_{j\neq k}^{\ell}m_{j}=M_{k}}\,\prod_{\stackrel{j=1}{j\neq k}}^{\ell}\frac{\left(n_{j}+m_{j}-1\right)!}{m_{j}!\left(n_{j}-1\right)!}\frac{1}{\left(a_{k}-a_{j}\right)^{n_{j}+m_{j}}}\label{eq:M_general}
\end{eqnarray}
 which correctly reduces to eq.~(\ref{eq:generating}) when all $n_{k}=1$.
As for the non-degenerate case, this function is regular at $y=0$,
and it is fact an entire function. The probability density can be
obtained with the same arguments used for the non-degenerate case
and the result is 
\begin{eqnarray}
P_{\mathcal{A}}\left(x\right) & = & \sum_{k=1}^{\ell}\sum_{M_{k}=0}^{n_{k}-1}\frac{\left(a_{k}-x\right)^{d+M_{k}-n_{k}-1}\mathrm{sign}\left(a_{k}-x\right)\left(-1\right)^{M_{k}}\left(d-1\right)!}{2\left(d+M_{k}-n_{k}-1\right)!\left(n_{k}-1-M_{k}\right)!}\times\nonumber \\
 &  & \sum_{\sum_{j\neq k}^{\ell}m_{j}=M_{k}}\,\prod_{\stackrel{j=1}{j\neq k}}^{\ell}\left(\begin{array}{c}
n_{j}+m_{j}-1\\
m_{j}
\end{array}\right)\frac{1}{\left(a_{k}-a_{j}\right)^{n_{j}+m_{j}}}\,.\label{eq:P_general}
\end{eqnarray}
 As a concrete example we will consider in some detail the case where
$A$ is a one-dimensional projector.

\subsection{Probability density for the random guess}

The case where $A$ is a one-dimensional projector is of some relevance
and we discuss it in some detail. Because of unitary invariance of
the measure we can fix the reference state to be $|1\rangle$ in some
basis, i.e.~$A=|1\rangle\langle1|$. In this case the quantity $\mathcal{A}$
that we are considering is the fidelity between a reference state
$|1\rangle$ and a random vector $|\psi\rangle$, i.e.~$\mathcal{A}=\left|\langle1|\psi\rangle\right|^{2}$
(normally written as $\mathcal{F}$). It is a standard textbook exercise
(see e.g.~\citep{preskill_course_2004} chapter two) to compute the
average random guess. Using eq.~(\ref{eq:average}) one gets $\overline{\mathcal{F}}=1/d$,
but what is the form of the probability distribution? Since $A$ is
a one-dimensional projector it has two eigenvalues $a_{1}=1$ and
$a_{2}=0$ with multiplicity $n_{1}=1$ and $n_{2}=d-1$ respectively.
Inserting these values in eq.~(\ref{eq:M_general}) we obtain for
the moment generating function

\begin{equation}
M_{\mathcal{F}}\left(y,\Omega\right)=\left(d-1\right)!\left\{ \frac{e^{y\left(1-\Omega\right)}}{y^{d-1}}-\sum_{m=0}^{d-2}\frac{e^{-y\Omega}}{y^{m+1}\left(d-2-m\right)!}\right\} \,.\label{eq:M_random-guess}
\end{equation}
 Once again, despite its appearance, this function is regular at $y=0$
and in fact entire. Rearranging the terms in the sum we can write
it as 
\begin{equation}
M_{\mathcal{F}}\left(y,\Omega\right)=e^{-\Omega y}\sum_{n=0}^{\infty}y^{n}\frac{\left(d-1\right)!}{\left(d+n-1\right)!}\,.
\end{equation}
 Using the above equation we readily obtain the moments as 
\begin{equation}
\overline{\mathcal{F}^{n}}=\frac{n!\left(d-1\right)!}{\left(d+n-1\right)!}\,,\label{eq:moments_guess}
\end{equation}
 which agrees with a result of Von Neumann \citep{neumann_beweis_1929,neumann_proof_2010}.
The probability density can be obtained either by Fourier transforming
eq.~(\ref{eq:M_random-guess}) or by setting $a_{1}=1,\, n_{1}=1$
and $a_{2}=0,\, n_{2}=d-1$ in eq.~(\ref{eq:P_general}). The sum
over $M_{2}$ becomes a binomial and the result is surprisingly simple
\begin{equation}
P_{\mathcal{F}}\left(x\right)=\left(d-1\right)\left(1-x\right)^{d-2}\1_{\left[0,1\right]}\left(x\right).\label{eq:P_random-guess}
\end{equation}
 Here we denoted by $\1_{\left[0,1\right]}\left(x\right)$ the indicator
function of the set $\left[0,1\right]$, i.e.~$\1_{\left[0,1\right]}\left(x\right)=\left[\mathrm{sign}\left(x\right)+\mathrm{sign}\left(1-x\right)\right]/2$.
Eq.~(\ref{eq:P_random-guess}) is the so-called beta distribution
with parameters $\alpha=1$ and $\beta=d-1$. In Bayesian statistics,
the beta distribution can be seen as the posterior probability of
the parameter $p$ of a binomial distribution after observing $\alpha-1$
successes (with probability of success given by $p$) and $\beta-1$
failures (with probability of failure $1-p$). The CDF is given by
a simple integration and reads 
\begin{equation}
\mathrm{cdf}_{\mathcal{F}}\left(x\right)=1-\left(1-x\right)^{d-1},\quad\mathrm{for}\;0\le x\le1\,,\label{eq:cdf_random-guess}
\end{equation}
 whereas $\mathrm{cdf}_{\mathcal{F}}\left(x\right)=0,\,(1)$ for $x<0$
($x>1$) respectively.

\section{Comparison with Levy's lemma}

A typical approach to gain information on the concentration properties
of a random variable $X$, is to compute the first few moments of
the variable and then use variations of Markov's or Chebyshev's inequalities
to obtain a bound on $\mathrm{Prob}\left(X-\overline{X}>\epsilon\right)$.
Alternatively in some cases one can use the Levy's lemma which typically
provides tighter bounds. Since we obtained the probability density
exactly we would like to compare the exact concentration prediction
with that obtained by Levy's bound.

Roughly speaking Levy's lemma states that, for a vector in a large-dimensional
hypersphere the probability that a (sufficiently smooth) function
is far from its average is exponentially small in the dimension of
the space. There are many versions of Levy's lemma involving either
the mean or the median or differing for the definition of being {}``far
from''. To be specific we use the following version \citep{ledoux_concentration_2001}:
for $x\in\mathbb{S}^{k}=\left\{ y\in\mathbb{R}^{k+1},\,\,\left\Vert y\right\Vert =1\right\} $,
and for a function $f$ with Lipschitz constant $\eta$ 
\begin{equation}
\mathrm{Prob}\left\{ f\left(x\right)-\langle f\rangle\ge\epsilon\right\} \le2\exp\left(-C_{1}\left(k+1\right)\epsilon^{2}/\eta^{2}\right)\label{eq:Levy}
\end{equation}
 with $C_{1}=\left(9\pi^{3}\ln2\right)^{-1}$. In our setting $|\psi\rangle$
lives in a $d$ dimensional complex space so $k=2d-1$.

We will compare the prediction of the Levy's lemma to our exact result
for two concrete examples of operators $A$.

\subsection{Random guess}

First we consider again the fidelity of the random guess, in which
case $A$ is a one-dimensional projector. As we mentioned previously,
the average fidelity is $\overline{\mathcal{F}}=1/d$, so to compute
the LHS of eq.~(\ref{eq:Levy}) it suffices to notice that $\mathrm{Prob}\left(\mathcal{F}>x\right)=1-\mathrm{cdf}_{\mathcal{F}}\left(x\right)$
and set $x=\overline{\mathcal{F}}+\epsilon=1/d+\epsilon$ in eq.~(\ref{eq:cdf_random-guess}).
Then

\begin{eqnarray}
\mathrm{Prob}\left(\mathcal{F}-\overline{\mathcal{F}}\ge\epsilon\right) & = & \left(1-\frac{1+\epsilon d}{d}\right)^{d-1}\nonumber \\
 & = & \frac{e^{-\left(d-1\right)/(d\left(1-\epsilon\right))}}{1-\epsilon}e^{\ln\left(1-\epsilon\right)d}\nonumber \\
 & \simeq & e^{-1}e^{-\epsilon d}\,,\label{eq:Levy-random-guess}
\end{eqnarray}
 where the last equation has been obtained in the limit of $\epsilon$
small and $d$ large. The Lipschitz constant of the function $\mathcal{A}\left(\phi\right)=\langle\phi|A|\phi\rangle$
has been calculated in \citep{popescu_foundations_2005} Appendix
A (see also \citep{popescu_entanglement_2006}) where it has been
shown that $\left|\mathcal{A}\left(\phi_{1}\right)-\mathcal{A}\left(\phi_{2}\right)\right|\le2\left\Vert A\right\Vert _{op}\left||\phi_{1}\rangle-|\phi_{2}\rangle\right|$
where $\left\Vert A\right\Vert _{op}$ is the operator norm of $A$
(maximum singular value). In our case $A=|1\rangle\langle1|$ so $\left\Vert A\right\Vert _{op}=1$
and we may take $\eta=2$ in eq.~(\ref{eq:Levy}) above. So the Levy's
lemma predicts 
\begin{equation}
\mathrm{Prob}\left(\mathcal{F}-\overline{\mathcal{F}}\ge\epsilon\right)\le2e^{-C\epsilon^{2}d}\label{eq:Levy_projector2}
\end{equation}
with constant $C=(18\pi^{3}\ln2)^{-1}\simeq1/387$. As we see comparing
with eq.~(\ref{eq:Levy-random-guess}) Levy's bound pays a very small
pre-factor in the exponential compared to the exact behavior and a
quadratic rather than a linear dependence on the error $\epsilon$.

\subsection{Number operator}

Let us now take an example from the non-degenerate case and take $A$
to be the number operator $\hat{N}$, i.e.~the operator with $a_{k}=k$
for $k=1,2,\ldots,d$. For this operator we have $\overline{\mathcal{A}}=\tr A/d=\left(d+1\right)/2$
and variance $\Delta\mathcal{A}^{2}=(d-1)/12$. Since the norm of
$\hat{N}$ grows linearly with $d$ we do not expect concentration
to take place in this case. Now

\begin{equation}
\mathrm{Prob}\left\{ \mathcal{A}-\overline{\mathcal{A}}\ge\epsilon\right\} =1-\mathrm{cdf}\left(\frac{d+1}{2}+\epsilon\right)=:B\left(d,\epsilon\right)
\end{equation}

We investigated numerically the function $B\left(d,\epsilon\right)$
using eq.~(\ref{eq:cdf}). Our results indicate that for large $d$
and small $\epsilon$, $B\left(d,\epsilon\right)\simeq(1/2)e^{-C\epsilon/\sqrt{d}}$
with constant $C\approx2.9$. This behavior is consistent with $P_{\mathcal{A}}(\overline{\mathcal{A}})\simeq C'/\sqrt{d}$
which has also been checked numerically using Eq.~(\ref{eq:P-density}).
The operator norm is $\left\Vert \hat{N}\right\Vert _{op}=d$ so Levy's
lemma in this case predicts the following bound 
\begin{equation}
B\left(d,\epsilon\right)\le2e^{-C\epsilon^{2}/d}
\end{equation}
 with same constant $C$ as in Eq.~(\ref{eq:Levy_projector2}). In
this case we see that the Levy lemma predicts a wrong scaling both
with respect to the error $\epsilon$ and the space dimension $d$.
In any case Levy lemma is sufficient to prove the following concentration
result for the rescaled operator $\hat{N}/d$. Namely that the random
variable $\langle\psi|\hat{N}|\psi\rangle/d$ tends in distribution
to a delta centered in $x=1/2$, as was first observed in \citep{bender_solvable_2005}.
This follows readily from the fact that the Lipschitz constant now
is $O(1)$ so that the right hand side of Eq.~(\ref{eq:Levy}) tends
to zero as $d\to\infty$.

\section{Central limit theorem and measure concentration}

If the (non-degenerate) self-adjoint operator $A$ has the following
spectral resolution $A=\sum_{k=}^{d}a_{k}|k\rangle\langle k|,$ our
random variable $\mathcal{A}(\psi)$ can be written as a weighted
sum of the random variables $X_{k}(\psi):=|\langle\psi|k\rangle|^{2}.$
In the limit of large Hilbert space dimension $d$ it is easy to see
that the $X_{k}$'s decouple. Indeed if $h\neq k$ one has 
\begin{eqnarray}
\overline{X_{k}X_{h}} & = & \overline{\langle\psi|k\rangle\langle k|\psi\rangle\langle\psi|h\rangle\langle h|\psi\rangle}=\overline{\tr\left[|\psi\rangle\langle\psi|^{\otimes\,2}|k\rangle\langle k|\otimes|h\rangle\langle h|\right]}\nonumber \\
 & = & \tr\left[\frac{1+P}{d(d+1)}|k\rangle\langle k|\otimes|h\rangle\langle h|\right]=\frac{1}{d(d+1)},
\end{eqnarray}
 while for $h=k$ $\overline{X_{k}^{2}}=2/d(d+1)$ from eq.~(\ref{eq:moments_guess})
with $n=2.$ Whence one obtains for the covariance: 
\begin{equation}
{\rm cov}(X_{h},X_{k}):=\overline{X_{k}X_{h}}-\overline{X_{k}}\,\overline{X_{h}}=\frac{1}{d(d+1)}-\frac{1}{d^{2}}=O(1/d^{3}),
\end{equation}
 for $h\neq k$ and $\mathrm{cov}(X_{k},X_{k})=O\left(2/d^{2}\right)$,
i.e.~the off-diagonal terms decay faster than the diagonal ones.
This fact is a consequence of a well known, general result. In physics
a manifestation of this phenomenon is the fact that the large-$N$
limit of $O\left(N\right)$ invariant field theories (such as the
non-linear sigma model) is a free, Gaussian, theory (see e.g.~\citep{auerbach_interacting_1994}).
Probabilists generally tend to see this result in the opposite direction,
namely: in the large $N$ limit, the Gaussian measure becomes highly
concentrated on a hyper-sphere of given radius. Given these considerations
it is natural to expect that the variable $\mathcal{A}$ satisfies
a central limit theorem (CLT), at least for a certain class of operators
$A$. By CLT we mean here that the rescaled random variable 
\begin{equation}
\mathcal{Z}:=\left(\mathcal{A}-\overline{\mathcal{A}}\right)/\sqrt{\kappa_{2}\left(\mathcal{A}\right)},
\end{equation}
 ($\kappa_{n}\left(\mathcal{A}\right)$ $n$-th cumulant) tends in
distribution to a Gaussian with zero mean and unit variance as $d\to\infty$.

\begin{figure}
\noindent \begin{centering}
\includegraphics[width=8cm]{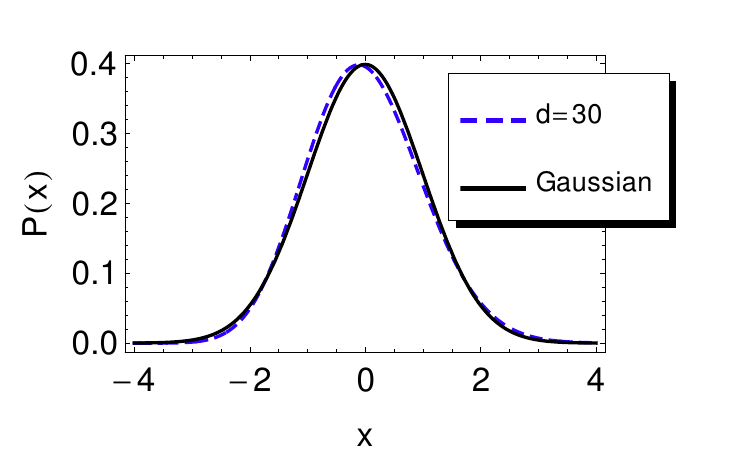}\includegraphics[width=8cm]{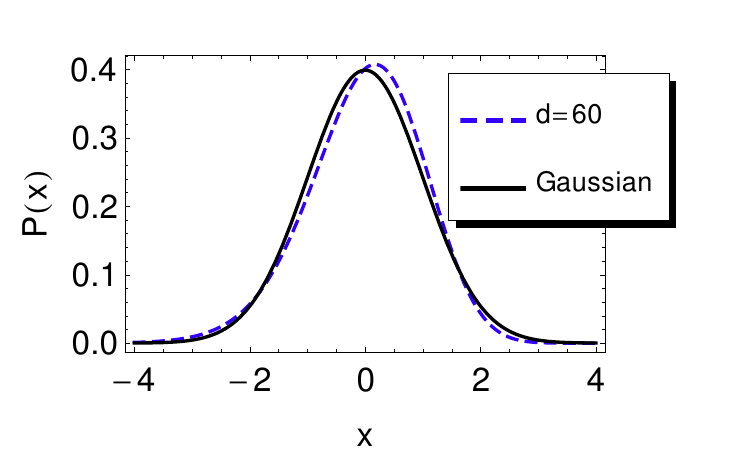} 
\par\end{centering}

\noindent \caption{Approach to Gaussian and central limit theorem for the rescaled variable
$\mathcal{Z}:=\left(\mathcal{A}-\overline{\mathcal{A}}\right)/\sqrt{\kappa_{2}\left(\mathcal{A}\right)}$.
Left panel: the operator $A$ has eigenvalues $a_{k}=k^{2}$. Right
panel: $a_{k}=\ln\left(k\right)$. The continuous line is the limiting
case given by a standard normal distribution with zero mean and unit
variance. \label{fig:CLT}}
\end{figure}

Instead of trying to give a complete characterization of the precise
conditions on $A$ for the CLT to apply we will content here with
a few examples (see however \citep{gallay_numerical_2012}). Consider
first the class of $A$ with $a_{k}=k^{\alpha}$ (with say $\alpha>0$).
Using formulae (\ref{eq:average}) and (\ref{eq:m2-m3}) we obtain,
at leading order $\kappa_{1}\left(\mathcal{A}\right)\propto d^{\alpha}$,
$\kappa_{2}\left(\mathcal{A}\right)\propto d^{2\alpha-1}$, and $\kappa_{3}\left(\mathcal{A}\right)\propto d^{3\alpha-2}$
from which we get $\kappa_{3}\left(\mathcal{Z}\right)\propto1/\sqrt{d}$,
i.e.~the third cumulant of $\mathcal{Z}$ vanishes for large $d$.
The approach to Gaussian for this case can be read off from figure
\ref{fig:CLT}. Let us give another, slightly more complicated example
for which $a_{k}=\log k$. In this case the first cumulant grows as
$\overline{\mathcal{A}}\propto\log d$. Proving that the third cumulant
of $\mathcal{Z}$ goes to zero is now less trivial. However Euler-Maclaurin
formula is sufficient to prove the following result valid at leading
order 
\begin{eqnarray}
\kappa_{2}\left(\mathcal{A}\right) & = & \frac{\sum_{k=1}^{d}\left(\ln\left(k\right)-\frac{\overline{x}}{d}\right)^{2}}{d\left(d+1\right)}=\frac{1}{d}+\mathrm{smaller\,\, terms}\\
\kappa_{3}\left(\mathcal{A}\right) & = & \frac{2\sum_{k=1}^{d}\left(\ln\left(k\right)-\frac{\overline{x}}{d}\right)^{3}}{d\left(d+1\right)\left(d+2\right)}=\frac{-4}{d^{2}}+\mathrm{smaller\,\, terms}
\end{eqnarray}
 where $\overline{x}=\sum_{k=1}^{d}\ln k=\ln\left(d!\right)$. Hence
we see that also in this case $\kappa_{3}\left(\mathcal{Z}\right)\propto1/\sqrt{d}$.
The approach to Gaussian is depicted in figure \ref{fig:CLT}.

In any case it should be clear that one has concentration for the
variable $\mathcal{A}$ as long as the eigenvalues of $A$ do not
grow too fast with the dimension $d$. To be more precise, using Levy's
lemma Eq.~(\ref{eq:Levy}) and the fact that the Lipschitz constant
of $\mathcal{A}$ is $\left\Vert A\right\Vert _{\mathrm{op}}$, we
see that whenever $\left\Vert A\right\Vert _{\mathrm{op}}=O(d^{1/2-\epsilon})$,
the variable $\mathcal{A}-\overline{\mathcal{A}}$ tends in distribution
to a delta centered around zero as $d\to\infty$. In the example considered
so far this happen for the case $a_{k}=k^{\alpha}$ for $\alpha<1/2$
and for $a_{k}=\ln k$. Consider now a physically relevant situation
where the Hilbert space is that of a many-body, multipartite system.
In this case particular importance is played by extensive operators
for which $A$ is a sum of local terms i.e.~$A=\sum_{x}A_{x}$. Here
the label $x$ runs over the volume $V$ of the system and the total
dimension $d$ is exponential in the volume, i.e.~$d=e^{\alpha V}$
with $\alpha$ positive constant. The operator norm for such extensive
operators is clearly $\left\Vert A\right\Vert _{\mathrm{op}}=O(\ln d)$
and so for Levy's lemma Eq.~(\ref{eq:Levy}) $\mathcal{A}$ converges
in distribution to a delta around its average $\overline{\mathcal{A}}$.
To be more precise fluctuations are exponentially small in the volume
up to possible logarithmic corrections.

\section{A note on {}``quantum microcanonical  equilibrium''\label{sec:A-note-on}}

We would like to draw here few comments on the so called quantum microcanonical
 equilibrium (QME) introduced in refs.~\citep{brody_quantum_1998,brody_microcanonical_2005,bender_solvable_2005,brody_quantum_2007,brody_quantum_2007-1}
(see also \citep{fine_typical_2009,fine_alternative_2010,ji_nonthermal_2011}).
In the QME setting the density of states of an isolated system with
Hamiltonian $H$ at energy $E$ is given precisely by $\Omega(E)=\overline{\delta\left(\langle\psi|H|\psi\rangle-E\right)}$.
The entropy is then defined in the usual way as $S=k_{B}\ln\Omega\left(E\right)$
(in the following we will set Boltzmann's constant $k_{B}$ equal
to one) which in turns allows to define the microcanonical  temperature
as $T^{-1}=\partial S/\partial E$. Quoting \citep{brody_microcanonical_2005}
{}``the advantage of the QME formulation over the traditional approach
is that the entropy is a continuous function of the energy. As a consequence,
thermodynamic functions {[}...{]} are well defined for finite quantum
systems''. Indeed, as we have seen, $\Omega(E)$ is a continuous
piecewise polynomial, and any thermodynamic function will be analytic
except for singular points located at the eigenvalues of $H$ where
the polynomials join. 

We will now try to answer a few simple questions related to QME, namely
what is the expected size of statistical fluctuations, whether QME
gives rise to extensive thermodynamic functions and briefly describe
the high and low temperature behavior. 

The statistical mechanical setting is that of many-body quantum systems
outlined in the end of the previous section. In particular the Hilbert
space dimension is exponential in the numbers of constituent ($d=e^{\alpha V}$).

According to the prescription of QME, the equilibrium average of an
observable $A$ is defined as $\langle A\rangle_{QME}:=\overline{\langle\psi|A|\psi\rangle}$
(see Eq.~(15) of \citep{brody_microcanonical_2005}). Because of
Eq.~(\ref{eq:average}) this average coincides with the standard
microcanonical  average $\langle A\rangle_{MC}=\tr\left(A\rho_{MC}\right)=\tr\left(A\right)/d$
and moreover it is extensive for extensive observables (i.e.~$A=\sum_{x}A_{x}$).
Let us then investigate the size of the fluctuations. Microcanonical
 (MC) fluctuations are given by $\Delta A_{MC}^{2}=\langle A^{2}\rangle_{MC}-\left[\langle A\rangle_{MC}\right]^{2}$
which acquire the familiar form $\Delta A_{MC}^{2}=\sum_{x,y}\left[\langle A_{x}A_{y}\rangle_{MC}-\langle A_{x}\rangle_{MC}\langle A_{y}\rangle_{MC}\right]$
for extensive observables. Fluctuations are then extensive whenever
the correlation $C_{x,y}=\left[\langle A_{x}A_{y}\rangle_{MC}-\langle A_{x}\rangle_{MC}\langle A_{y}\rangle_{MC}\right]$
is a sufficiently fast decaying function of $\left|x-y\right|$ at
large separations ($C_{x,y}\sim|x-y|^{-D-\epsilon}$, with $D$ spatial
dimension and $\epsilon>0$ is sufficient). The second moment of an
observable in the QME framework should be computed according to $\langle A^{2}\rangle_{QME}=\overline{\langle\psi|A|\psi\rangle^{2}}$.
This is particularly clear when $A$ is the Hamiltonian $H$ itself
since the moments of the energy are given by $\int E^{n}\Omega(E)dE=\overline{\langle\psi|H|\psi\rangle^{n}}$.
Using Eq.~(\ref{eq:m2-m3}) this leads to fluctuations given by $\Delta A_{QME}^{2}=\Delta A_{MC}^{2}/(d+1)$.
 Contrary to the standard case, fluctuations are exponentially small
in the volume $V$ (remind $d=\exp\alpha V$). This is just another
manifestation of the concentration phenomenon that we have discussed
for the probability distribution $P_{\mathcal{A}}(a)$. At equilibrium,
observables in the QME framework are concentrated around the same
values as in the traditional mirocanonical  setting but with much
smaller variances. It appears that the reason of the departure of
the QME from the traditional MC formulation, is that the density of
states at energy $E$ is given by the volume of those states whose
\emph{expectation value} is $E$. Whereas in the traditional approach
$\Omega_{MC}(E)$ is the volume of the states whose energy is \emph{exactly}
$E$ (i.e.~the Hamiltonian eigenstates). Indeed, while $\Omega_{QME}(E)=\overline{\delta\left(\langle\psi|H|\psi\rangle-E\right)}$,
the traditional density of states can be written as $\Omega_{MC}(E):=\overline{\langle\psi|\delta\left(H-E\right)|\psi\rangle}=\langle\delta(H-E)\rangle_{MC}$. 

We turn now to the canonical  ensemble. The canonical QME partition
function at inverse temperature $\beta$ is given by \citep{brody_quantum_1998,brody_microcanonical_2005}
(for simplicity we write $\langle H\rangle$ in place of $\langle\psi|H|\psi\rangle$
now on) 
\begin{equation}
Z_{QME}(\beta)=d\overline{e^{-\beta\langle H\rangle}}
\end{equation}
and has been computed in Sec.~\ref{sec:General-case} %
\footnote{The factor $d$ is needed because the partition function must be normalized
to the dimension of the space.%
}. In particular we have shown that, for finite systems, $Z_{QME}(\beta)$
is an analytic function of $\beta$ over the whole complex plane.
The free energy is given as usual by $F_{QC}(\beta)=-(1/\beta)\ln Z_{QME}$.
Using Jensen's inequality repeatedly (in particular $\langle e^{-\beta X}\rangle\ge e^{-\beta\langle X\rangle}$
for any measure $\langle\cdot\rangle$) we obtain 
\begin{equation}
\frac{1}{d}\tr e^{-\beta H}=\overline{\langle e^{-\beta H}\rangle}\ge\overline{e^{-\beta\langle H\rangle}}\ge e^{-\beta\overline{\langle H\rangle}}=e^{-\beta\tr H/d}.
\end{equation}
Multiplying the above equation by $d$ and taking the logarithm, we
draw the comforting result that $F_{QMC}(\beta)$ is of the order
of the volume (whenever the standard free energy is). Indeed one obtains
\begin{equation}
F_{C}(\beta)\le F_{QME}\left(\beta\right)\le Ve_{MC}-\frac{\alpha}{\beta}V\label{eq:F_bounds}
\end{equation}
where we set $Ve_{MC}=\tr H/d$ and wrote $F_{C}(\beta)$ for the
extensive canonical  free energy. Moreover, from the previous discussion,
we know that $\beta F_{QME}(\beta)$ is an analytic function ($Z_{QME}(\beta)$
is) for finite systems. 

Let us now investigate basic properties of the QME free energy in
the high and low temperature limit. 

Consider first the high temperature regime around $\beta=0$. In this
limit the series expansion of the free energy is given by the cumulants
of the corresponding distribution. To be more precise we have
\begin{equation}
\ln Z_{QME}(\beta)=\ln d+\sum_{n=1}^{\infty}\frac{(-\beta)^{n}}{n!}\kappa_{n}^{QME}\,.
\end{equation}
In the above equation we indicated $\kappa_{n}^{QME}$ for the cumulants
in the QME case, which can be obtained essentially taking the logarithm
of Eq.~(\ref{eq:chi-series}) and expanding around $\lambda=0$.
At $\beta=0$ the cumulants of the QME distribution $\kappa_{n}^{QME}$
can be related to the moments of the standard microcanonical distribution
(see e.g.~Eqns.~(\ref{eq:m2-m3}) and (\ref{eq:m4})) which in turn
can be expressed as cumulants. Moreover we can assume, as it is often
the case, that such standard microcanonical cumulants are all extensive,
i.e.~$\kappa_{n}^{MC}=O(V)$, in other words the system has no phase
transition at $\beta=0$ (in the standard framework). As we have seen
previously the first cumulants coincide: $\kappa_{1}^{QME}=\kappa_{1}^{MC}=\langle H\rangle_{MC}$.
For the second cumulant we obtained $\kappa_{2}^{QME}=\Delta^{2}H_{MC}/(d+1)$,
which vanishes as $V\to\infty$. Expressing the cumulants through
the moments and using Eq.~(\ref{eq:m2-m3}), we can obtain the third
cumulant as
\begin{eqnarray}
\kappa_{3}^{QME} & = & \frac{2\kappa_{3}^{MC}}{(d+1)(d+2)}\label{eq:k3_QME}
\end{eqnarray}
which vanishes exponentially in the volume as $V\to\infty$. With
some extra work we can get the following expression for the fourth
cumulant:
\begin{equation}
\kappa_{4}^{QME}=6\frac{\kappa_{4}^{MC}(d+1)+\left[\kappa_{2}^{MC}\right]^{2}d}{(d+1)^{2}(d+2)(d+3)}\,.\label{eq:k4_QME}
\end{equation}
Again, given the extensivity assumption assumption of MC cumulants,
this cumulant vanishes exponentially in the volume. Indeed, as we
discussed in the previous section, for extensive observables Levy's
lemma guarantees that the variable $\langle\psi|H|\psi\rangle$ converges
in distribution to a delta around its average. This implies that the
behavior observed in Eqns.~(\ref{eq:k3_QME}) and (\ref{eq:k4_QME})
continues to all orders and all the cumulants except the first converge
to zero, i.e.~$\kappa_{n}^{QME}\to0$ for $n>1$. 

This implies that in the high temperature limit the upper bound in
Eq.~(\ref{eq:F_bounds}) is saturated and the free energy density
assumes the form $f_{QME}(\beta):=\lim_{V\to\infty}F_{QME}(\beta)/V=e_{MC}-\alpha/\beta$,
in a neighborhood of $\beta=0$. Actually one has the stronger result
$\lim_{V\to\infty}\ln Z_{QME}(\beta)+e_{MC}V\beta=0$. 

Since $\ln Z_{QME}(\beta)$ is not a straight line in the low temperature
regime (see below), this imply some sort of temperature-driven phase
transition irrespective of the model under consideration. 

Let us now look at the zero temperature limit ($\beta\to\infty$).
In this case, exploiting Eq.~(\ref{eq:chi}) we get, at leading order,
\begin{equation}
\ln Z_{QME}(\beta)\simeq-\beta E_{0}+\ln(d!C_{0})-(d-1)\ln\left(\beta\right)+\frac{C_{1}}{C_{0}}e^{-\beta\Delta}\,,
\end{equation}
with $E_{0}$ the ground state energy, $C_{k}=\prod_{j\neq k}\left(E_{k}-E_{j}\right)$,
and smallest gap $\Delta$. This slow approach to zero temperature
has to be contrasted with the familiar exponential approach $\ln Z_{C}(\beta)\simeq-\beta E_{0}+e^{-\beta\Delta}$.
The mean energy in this limit has the form 
\begin{equation}
\langle H\rangle_{QME}\simeq E_{0}+\frac{d-1}{\beta}+\frac{C_{1}}{C_{0}}\Delta e^{-\beta\Delta}.
\end{equation}
The ground state energy is not reached exponentially fast (as it happens
normally for gapped systems with $\Delta\neq0$) but algebraically
in the temperature. Moreover, consistently with ref.~\citep{brody_quantum_1998},
this implies that the specific heat approaches a (large) constant
$C_{V}=\partial E/PT\to(d-1)$ at zero temperature irrespective of
the presence of a gap in the spectrum.

\section{Conclusions}

In this paper we considered the quantum expectation value of an operator
$A$ with respect to a pure state $|\psi\rangle$. We computed exactly
the probability density of the expectation value when $|\psi\rangle$
is drawn from the space of pure states according to the unique (Haar-induced)
unitarily invariant measure. Generically, the resulting probability
distribution is a compactly supported, piecewise polynomial function.
We used the exact result to test the tightness of the concentration
bounds obtained by Levy's lemma for a couple of particular cases,
namely for $A$ one-dimensional projector and for $A=\hat{N}$ the
number operator. Levy's lemma reproduced a qualitatively correct scaling
behavior with the dimension of the Hilbert, although with very small
pre-factor as compared to the exact ones. The quadratic scaling with
the error $\epsilon$ predicted by Levy's lemma was seen to reduce
to linear scaling for sufficiently small $\epsilon$ in the cases
studied. We have also noticed that quantum expectation of $A$ can
be regarded as a linear combination of random variables that decouple
in the limit of large Hilbert space dimension. Here we limited ourselves
to discuss a couple of examples of quantum operators whose expectation
value fulfill a central limit type of result in such a limit i.e.,
a properly rescaled expectation becomes normally distributed. We have
also applied some of our results to study the size of fluctuations
and the extensivity of thermodynamic functions of an alternative approach
to equilibrium developed in \citep{brody_quantum_1998,brody_microcanonical_2005}
that goes under the name of {}``quantum microcanonical  equilibrium''. 

Before concluding it is important to mention that the Haar-induced
measure over quantum pure states is well-known to be unphysical in
different ways. In fact, sampling this measure with local quantum
gates requires exponentially long random circuits \citep{emerson_pseudo-random_2003}.
Also, Haar typical quantum states are nearly maximally entangled \citep{hayden_aspects_2006},
while low energy eigenstates of local quantum Hamiltonian fulfill
area laws \citep{eisert_colloquium:_2010,hamma_ground_2005,hamma_bipartite_2005}
i.e., they have low entanglement. In view of these remarks one may
question the physical relevance of the results presented in this paper
and look for more constrained prior measures over the $|\psi\rangle$'s.
For example, in view of applications to foundations of statistical
mechanics of closed systems, it would be interesting to generalize
our results to {}``sections of constant energy'' where one draws
pure states uniformly under some constraint of the form $\langle\psi|H|\psi\rangle=E$
as e.g.~in the spirit of \citep{muller_concentration_2011}. 

A more ambitious goal would be to determine the distribution of expectations
restricted to a set of \textquotedbl{}physical\textquotedbl{} states
endowed with some {}``natural\textquotedbl{} measure. Instances of
those ensembles are given in \citep{hamma_quantum_2011,garnerone_typicality_2010,garnerone_statistical_2010}
where local random quantum circuits and matrix product states respectively
have been considered. In these cases the lack of the maximal unitary
invariance of the Haar measure represents the major obstruction one
has to overcome.

\paragraph*{Acknowledgement}

The authors would like to thank Karol \.{Z}yczkowski for bringing
to their attention references \citep{dunkl_numerical_2011,dunkl_numerical_2011-1,puchala_restricted_2012,gallay_numerical_2012}.

This research is partially supported by the ARO MURI grant W911NF-11-1-0268
and NSF grants No.~PHY-969969 and No.~PHY-803304.

\appendix

\section{Some useful identities\label{sec:Some-useful-identities}}

Here we want to prove some identities which provide several important
relations for the coefficients in Eq.~(\ref{eq:chi}) and eq.~(\ref{eq:P-density}).
We begin by considering the following function for $y,\Omega$ real
$M\left(y,\Omega\right):=\overline{e^{\left(\mathcal{A}-\Omega\right)y}}$
which is somehow the moment generating function with a shift. The
corresponding Gaussian integral is well defined when $y\left(A-\Omega\1\right)<0$.
To satisfy this condition we assume $\Omega>\max\{a_{j}\}$ and $y>0$.
The Gaussian integral converges and gives 
\begin{equation}
M\left(y,\Omega\right)=\frac{\left(d-1\right)!}{\left(-i\right)^{d}}\int\frac{dr}{2\pi}\frac{e^{-ir}}{\prod_{j}\left(r-r_{j}\right)},\quad r_{j}=iy\left(a_{j}-\Omega\right)
\end{equation}
 The integral can be evaluated again with complex integration. The
contour integral must be closed in the lower half plane, all the poles
are on the negative imaginary axis and we get 
\begin{eqnarray}
M\left(y,\Omega\right) & = & \frac{\left(d-1\right)!}{\left(-i\right)^{d-1}}\sum_{k=1}^{d}\frac{e^{-ir_{k}}}{\prod_{j\neq k}\left(r_{k}-r_{j}\right)}\nonumber \\
 & = & \frac{\left(d-1\right)!}{y^{d-1}}\sum_{k=1}^{d}\frac{e^{y\left(a_{k}-\Omega\right)}}{\prod_{j\neq k}\left(a_{k}-a_{j}\right)}\label{eq:generating}
\end{eqnarray}
 The same quantity at leading order in $y$ can be computed by first
expanding the integral around $y=0$ and the integrating. We obtain
then 
\begin{equation}
M\left(y,\Omega\right)=1+O\left(y\right)\label{eq:M-series}
\end{equation}
 Equating eqns.~(\ref{eq:generating}) and (\ref{eq:M-series}) term
by term we arrive at 
\begin{equation}
\sum_{k=1}^{d}\frac{\left(a_{k}-\Omega\right)^{n}}{\prod_{j\neq k}\left(a_{k}-a_{j}\right)}=\left\{ \begin{array}{cl}
0 & 0\le n\le d-2\\
1 & n=d-1
\end{array}\right.\,.\label{eq:nano}
\end{equation}
 These equations have been obtained assuming $\Omega>\max\{a_{j}\}$,
but since they are analytic in $\Omega$ they must be true for all
complex $\Omega$. In particular, setting $\Omega=0$ in (\ref{eq:nano})
we deduce that $\chi\left(\lambda\right)$ is regular at $\lambda=0$,
and in fact analytic and we obtain eq.~(\ref{eq:chi-series}). Considering
instead $\Omega>\max\{a_{k}\}$ (resp.~$\Omega<\min\{a_{k}\}$) and
applying the result to eq.~(\ref{eq:P-density}) we obtain that $P_{\mathcal{A}}\left(x\right)=0$
for $x>\max\{a_{k}\}$ (resp.~$x<\min\{a_{k}\}$).

\section{Degenerate case\label{sec:Degenerate-case}}

We provide here some additional steps needed to obtain eq.~(\ref{eq:M_random-guess}).
In practice we need to write down the differentiation in eq.~(\ref{eq:residue}).
The first step is the following: 
\begin{equation}
\partial_{r}^{\left(n_{k}-1\right)}\left[e^{-ir}g\left(r\right)\right]=\sum_{M=0}^{n_{k}-1}\left(\begin{array}{c}
n_{k}-1\\
M
\end{array}\right)\left(-i\right)^{n_{k}-1-M}e^{-ir}\partial_{r}^{M}g\left(r\right)
\end{equation}
 Then we need the multinomial formula 
\begin{equation}
\partial_{r}^{M}\left[\prod_{j=1}^{\ell}g_{i}\left(r\right)\right]=\sum_{m_{1}=0}^{M}\cdots\sum_{m_{\ell}=0}^{M}\,\delta_{M,\sum_{j=1}^{\ell}m_{j}}\, M!\prod_{j=1}^{\ell}\frac{g_{j}^{\left(m_{j}\right)}\left(r\right)}{m_{j}!}
\end{equation}
 As customary in physics we write the multiple sum as 
\begin{equation}
\sum_{m_{1}=0}^{M}\cdots\sum_{m_{\ell}=0}^{M}\,\delta_{M,\sum_{j=1}^{\ell}m_{j}}\,=\sum_{\sum_{j=1}^{\ell}m_{j}=M}.
\end{equation}
 Note that when applying this to eq.~(\ref{eq:residue}) we miss
the term with $j=k$ so we have $\ell-1$ (constrained) sums over
$m_{i}$. The final bit is 
\begin{equation}
\partial_{r}^{m_{j}}\left[\left(r-r_{j}\right)^{-n_{j}}\right]=\left(-1\right)^{m_{j}}\frac{\left(n_{j}+m_{j}-1\right)!}{\left(n_{j}-1\right)!}\frac{1}{\left(r-r_{j}\right)^{n_{j}+m_{j}}}\,.
\end{equation}
 Putting things together, the derivative reads 
\begin{eqnarray}
 &  & \partial_{r}^{\left(n_{k}-1\right)}\left[\frac{e^{-ir}}{\prod_{j\neq k}^{\ell}\left(r-r_{j}\right)^{n_{j}}}\right]=\\
 &  & \sum_{M=0}^{n_{k}-1}\left(\begin{array}{c}
n_{k}-1\\
M
\end{array}\right)\left(-i\right)^{n_{k}-1-M}e^{-ir}\times\\
 & \times & \sum_{\sum_{j\neq k}^{N}m_{j}=M}M!\prod_{j\neq k}\left(-1\right)^{m_{j}}\frac{\left(n_{j}+m_{j}-1\right)!}{m_{j}!\left(n_{j}-1\right)!}\frac{1}{\left(r-r_{j}\right)^{n_{j}+m_{j}}}
\end{eqnarray}

Going back we get 
\begin{eqnarray}
M\left(y,\Omega\right) & = & \frac{\left(d-1\right)!}{\left(-i\right)^{d-1}}\sum_{k}\frac{1}{\left(n_{k}-1\right)!}\times\\
 &  & \sum_{M_{k}=0}^{n_{k}-1}\left(\begin{array}{c}
n_{k}-1\\
M_{k}
\end{array}\right)\left(-i\right)^{n_{k}-1-M_{k}}e^{-ir_{k}}\times\\
 & \times & \sum_{\sum_{j\neq k}^{N}m_{j}=M_{k}}M_{k}!\prod_{j\neq k}\frac{\left(n_{j}+m_{j}-1\right)!}{m_{j}!\left(n_{j}-1\right)!}\frac{\left(-1\right)^{m_{j}}}{\left(r_{k}-r_{j}\right)^{n_{j}+m_{j}}}
\end{eqnarray}
 Correctly all $i$ factors cancel out and we are left with 
\begin{eqnarray}
M\left(y,\Omega\right) & = & \left(d-1\right)!\sum_{k=1}^{\ell}\sum_{M_{k}=0}^{n_{k}-1}\frac{\left(-1\right)^{M_{k}}e^{y\left(a_{k}-\Omega\right)}}{y^{d+M_{k}-n_{k}}\left(n_{k}-1-M_{k}\right)!}\,\beta_{k}\left(M_{k}\right)\label{eq:f_general-1}\\
\beta_{k}\left(M_{k}\right) & = & \sum_{\sum_{j\neq k}^{\ell}m_{j}=M_{k}}\,\prod_{\stackrel{j=1}{j\neq k}}^{\ell}\frac{\left(n_{j}+m_{j}-1\right)!}{m_{j}!\left(n_{j}-1\right)!}\frac{1}{\left(a_{k}-a_{j}\right)^{n_{j}+m_{j}}}
\end{eqnarray}
 which correctly reduces to eq.~(\ref{eq:generating}) when all $n_{k}=1$.
On the other hand we still have $M\left(y,\Omega\right)=1+\frac{y}{d}\tr\left(A-\Omega\right)+O\left(y^{2}\right)$
which can be used to show directly that $M\left(y,\Omega\right)$
is regular at $y=0$ and hence analytic in the whole complex plane.

\bibliographystyle{unsrt}
\bibliography{Haar}

\end{document}